# Engineering the Quantum Foam

Reginald T. Cahill
*School of Chemistry, Physics and Earth Sciences,
Flinders University, GPO Box 2100, Adelaide 5001, Australia*
Reg.Cahill@flinders.edu.au

______________________________________________________________

## ABSTRACT

In 1990 Alcubierre, within the General Relativity model for space-time, proposed a scenario for 'warp drive' faster than light travel, in which objects would achieve such speeds by actually being stationary within a bubble of space which itself was moving through space, the idea being that the speed of the bubble was not itself limited by the speed of light. However that scenario required exotic matter to stabilise the boundary of the bubble. Here that proposal is re-examined within the context of the new modelling of space in which space is a quantum system, *viz* a quantum foam, with on-going classicalisation. This model has lead to the resolution of a number of longstanding problems, including a dynamical explanation for the so-called `dark matter' effect. It has also given the first evidence of quantum gravity effects, as experimental data has shown that a new dimensionless constant characterising the self-interaction of space is the fine structure constant. The studies here begin the task of examining to what extent the new spatial self-interaction dynamics can play a role in stabilising the boundary without exotic matter, and whether the boundary stabilisation dynamics can be engineered; this would amount to quantum gravity engineering.

## 1 Introduction

The modelling of space within physics has been an enormously challenging task dating back in the modern era to Galileo, mainly because it has proven very difficult, both conceptually and experimentally, to get a 'handle' on the phenomenon of space. Even then some major experimental bungles [1] have only recently been uncovered in 2002 [2,3], that lead to profoundly misleading concepts that formed the foundations of 20[th] century physics. Galileo and then Newton modelled space as an unchanging Euclidean 3-geometry, in which there was in principle no limit to the speed of objects. Einstein, building upon the theoretical work of Lorentz and the experimental work of Michelson and Morley [1], modified Lorentzian relativity to what is now known as Einsteinian relativity. The key concept here is the amalgamation of the geometrical model of space and time into, ultimately, a curved 4-dimensional pseudo-Riemannian spacetime manifold, giving General Relativity (GR), where the curvature models the phenomenon of gravity, unlike the Newtonian modelling of gravity which involved an acceleration vector field residing in the 3-space.

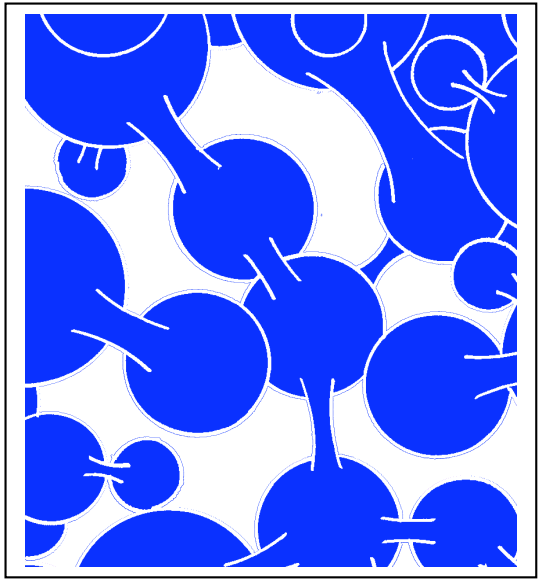

Fig.1 Artistic sketch of the quantum foam network that is space at its deepest level as emerges in the information theoretic *Process Physics*. Numerical studies have shown that the connectivity of this network is embeddable in a three-dimensional space, which is why this network is identified as that phenomenon which we know of as space. The blobs are *gebits* which to a first approximation are $S^3$ hyperspheres. These are linked via homotopic mappings. This whole connectivity pattern is fractal, in that any one of the gebits has this form for its internal structure.

Experimental evidence has resulted in the wide acceptance within physics of the curved spacetime model. However only recently [4,5,6,7] has it become clear that in those cases where the curved spacetime was experimentally and observationally successfully tested, the spacetime formalism turns out to have been nothing more that a 'flowing-space' system whose fundamental dynamical degree of freedom is a velocity field. Furthermore numerous experiments over the last 100 years or so have repeatedly and consistently reported the detection of this velocity field [3]. In particular any time-dependence and/or spatial inhomogeneity of this velocity field gives rise to the phenomenon we know of as gravity. At its deepest level this 'flowing space' is a classical description of a processing *quantum foam* [6,7,8].

Within both GR and the new theory of space the speed of light is the limiting speed of matter through space. However Alcubierre [9] has pointed out that this speed limit may be effectively bypassed if the matter is at rest within a bubble of space which itself is moving through space at greater than the speed of light. Elegant as this very non-Newtonian effect is, this proposal failed within GR because it required the presence of exotic matter to dynamically stabilise the boundary, as we later show, namely matter with essentially a 'negative mass'. Here we begin the task of examining how far the new spatial self-interaction dynamics can go in removing the need for such exotic matter, and whether any residual requirements for boundary stabilisation can be achieved by means of innovative engineering, that is by essentially 'engineering the quantum foam'.

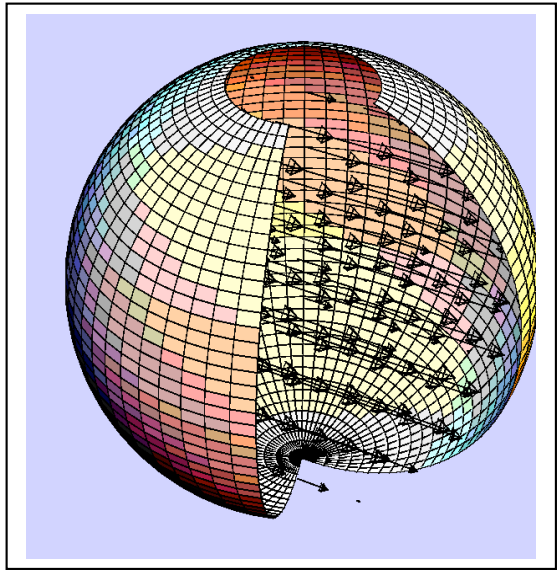

Fig.2 Velocity field for the propagating bubble given in (19)-(20). The velocity field is relative to a frame of reference in which the velocity field is zero outside of the bubble. Inside and outside of the bubble the flow satisfies both (2) and (5). The stability of this propagating bubble of space is then determined by the surface dynamics. The dynamics in (2), which is equivalent to GR, requires exotic matter at the boundary, as shown in Fig.4. However the key insight reported here is that this exotic matter may be replaced by the more complex self-interaction dynamics of the new theory of space, as given in (5). This produces an effective matter density as shown in Fig.5

## 2 Quantum Foam and its Flow Dynamics

The new theory of space arises within an information-theoretic modelling of reality, known as *Process Physics* [5,7,8]; essentially space and matter are emergent phenomena within a self-organising fractal pattern system, where both space and matter appear to be described by the new Quantum Homotopic Field Theory (QHFT). Therein space, at the deepest level, has the form represented with much artistic licence in Fig.1, where the fractal patterns form embedded and /or linked *gebits*, where the linking characteristics show that, at a coarse-grained level, there is an effective embeddability of the quantum-foam pattern structure within an abstract, ie not real, curved three-dimensional space. Because of the self-organising and processing of this quantum foam it essentially has differential motion, ie some regions 'move' relative to other regions. Of course this quantum foam is not embedded in any real background geometrical space. At the coarse-grained classical level this differential flow would be modelled by a velocity field, with the velocity field defined by reference to an arbitrary 'observer' or, more impersonally, to an arbitrary frame of reference. Covariance arguments quickly lead to the necessary minimal structure for the dynamics that must determine this velocity field; the change of frame of reference must not change the descriptive formalism, as the choice of reference frame is arbitrary. Differential flow is minimally described by an acceleration field, and to be Galilean covariant it must have the form

$$\mathbf{g} = \frac{d\mathbf{v}}{dt} \equiv \frac{\partial \mathbf{v}}{\partial t} + (\mathbf{v}.\nabla)\mathbf{v} \qquad (1)$$

which has been long-known as the Euler acceleration, first discovered in the context of classical fluids, ie matter flowing though a space. Matter effectively acts as a sink for the flow of the quantum foam, and the simplest non-relativistic description of matter is as a scalar density, and to relate the flow dynamics in (1) to this density we must have

$$\nabla.\left(\frac{\partial \mathbf{v}}{\partial t}+(\mathbf{v}.\nabla)\mathbf{v}\right)=-4\pi G\rho(r,t) \qquad (2)$$

where $G$ turns out to be the Newtonian gravitational constant. Outside of a spherical mass $M$ (2) has a time-independent radial in-flow solution (an alternate radial out-flow solution is unstable at the micro-level).

$$\mathbf{v}(\mathbf{r})=-\sqrt{\frac{2GM}{r}}\hat{\mathbf{r}} \qquad (3)$$

which using (1) then gives

$$\mathbf{g}(\mathbf{r})=-\frac{GM}{r^2}\hat{\mathbf{r}} \qquad (4)$$

So covariance requires this Newtonian inverse square law, at least minimally. Eqn.(2) may be generalised with the covariance maintained by including the next simplest structure

$$\frac{\partial}{\partial t}(\nabla.\mathbf{v})+\nabla.((\mathbf{v}.\nabla)\mathbf{v})+C(\mathbf{v})=-4\pi G\rho \qquad (5)$$

where

$$C(\mathbf{v})=\frac{\alpha}{8}\left((trD)^2-tr(D^2)\right); \quad D_{ij}=\frac{1}{2}\left(\frac{\partial v_i}{\partial x_j}+\frac{\partial v_j}{\partial x_i}\right) \qquad (6)$$

Eqn.(5) also has solution (3), and so acceleration (4), external to a spherical mass, and so in the solar system, with this mass being the sun, (5) is consistent with Kepler's laws for planetary motion. However (2), which is exactly Newtonian gravity within the velocity field formalism, differs from (5) within a spherically symmetric mass, and the difference manifests as the bore-hole $g$ anomaly. Fitting that data lead to the major discovery [10,11] that $\alpha$ has the same numerical value as the fine structure constant, to within experimental errors. Eqn.(5) has wave solutions as well as black hole solutions, and has explained the spiral galaxy rotation curve anomaly, and correctly predicted the mass of dynamically mandated black holes within globular clusters. The $C(\mathbf{v})$ term may be written on the RHS of (2) as an additional effective matter density.

$$\rho_{DM}=\frac{\alpha}{32\pi G}((trD)^2-tr(D^2)) \qquad (7)$$

which plays the role of the 'dark matter' (DM) effect in various systems, particularly spiral galaxies. Of course $\rho_{DM}(r,t)$ is not necessarily positive definite, and so in some circumstances this purely spatial self-interaction dynamics can mimic exotic 'negative mass' effects. Eqns.(2) and (5) can only be solved if $\mathbf{v}(\mathbf{r},t)$ has zero vorticity; $\nabla\times\mathbf{v}(\mathbf{r},t)=\mathbf{0}$. For non-zero vorticity more general arguments show that 2$^{nd}$-rank tensor flow equations may be constructed [5,8], and which at the simplest level introduce the vorticity induced by moving matter according to

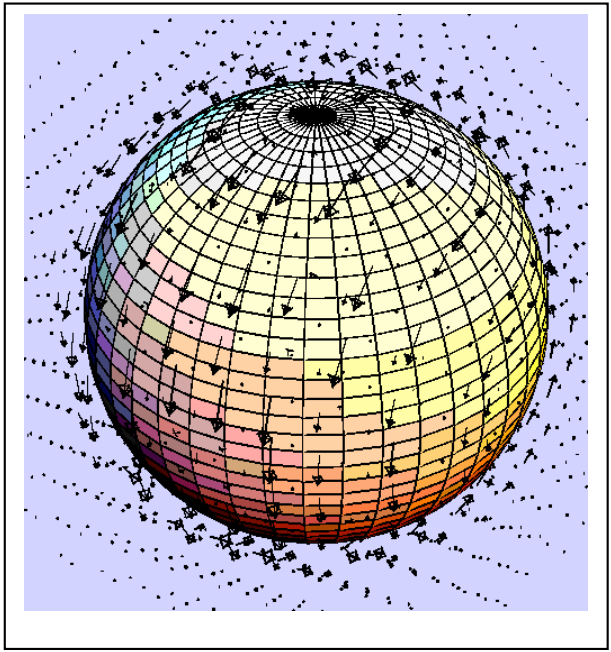

Fig.3 Shows the vorticity field $(\nabla \times \mathbf{v})$ of the velocity field in Fig.2. The bubble is moving towards the right. This vorticity occurs in the boundary layer of the propagating bubble of space, as specified in (19)-(20). Such vorticity must be produced by moving matter, as shown in (8), or perhaps by electromagnetic fields

$$\nabla \times (\nabla \times \mathbf{v}) = \frac{8\pi G \rho}{c^2} \mathbf{v}_R \qquad (8)$$

where $\mathbf{v}_R$ is the velocity of the matter relative to the 3-space. The form of the RHS of (8) has been confirmed to within 10% in [12]. The Gravity Probe B satellite gyroscope experiment is designed to study the vorticity from (8) induced by the rotation of the earth, but as well the new space theory implies that the linear motion of the earth will induce an additional component to the vorticity [13]  As we shall see the Alcubierre bubble of space necessarily involves non-zero vorticity at the boundary, and so involves the spatial dynamics in (8). The full flow theory of space briefly outlined above accounts for all the effects that supposedly *confirmed* GR, but goes further in explaining other various other key effects which GR is unable to account for, the most significant being the 'dark matter' effect.   This is easily seen because the 'dark matter' effect in (5) involves $\alpha$ as a second gravitational constant, whereas GR, like Newtonian gravity, involves only $G$.

## 4 General Relativity as Spatial Flow Dynamics

Because the Alcubierre proposal was originally formulated within the GR spacetime geometrical formalism, and because the current interpretation of that formalism makes no mention of an underlying spatial flow, it is insightful to briefly review the dynamical content of GR, in those cases where experiment, it is argued, has confirmed the formalism.

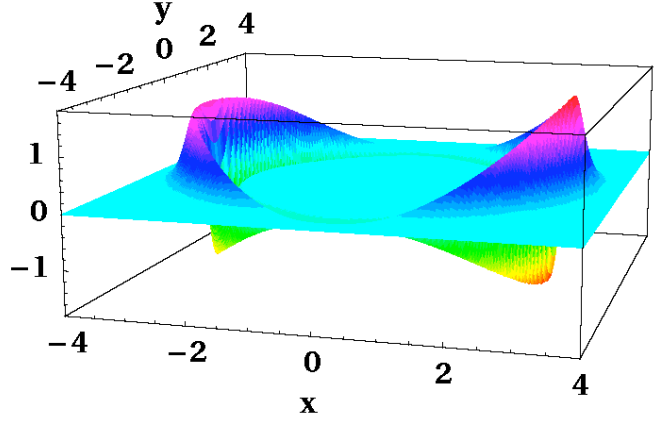

Fig.4 Shows the matter density on the RHS of (2) required in order that the propagating bubble in (19)-(20) satisfies (2), which is equivalent to GR. The plot shows the density on a plane passing through the centre of the bubble. The matter density, which resides in the boundary layer of the bubble, must be exotic, for we see that it must be negative in some regions.

From the beginning the equivalence principle, which goes back to Galileo, has been recognised as a key feature of gravity, namely that the accelerations of small test objects are independent of their mass. It has been argued, incorrectly as it now turns out, that this principle requires a metric theory of gravity, where the metric $g_{\mu\upsilon}(x)$ specifies the intrinsic structure of a spacetime construct according to the *interval*

$$d\tau^2 = g_{\mu\upsilon}(x)dx^\mu dx^\upsilon \qquad (9)$$

In GR the metric is the fundamental phenomenon; it characterises the curved spacetime. For time-like intervals, $d\tau^2 > 0$, and $d\tau$ is the elapsed time according to a co-moving clock. Then (9) has the integral form

$$\tau = \int \sqrt{g_{\mu\upsilon}(x)\frac{dx^\mu}{dt}\frac{dx^\upsilon}{dt}}\,dt \qquad (10)$$

The trajectory of a low-mass test object is determined by extremising $\tau$: $\delta\tau/\delta x^\mu = 0$, for a given $g_{\mu\upsilon}(x)$, which gives, in terms of the affine connection $\Gamma^\lambda_{\mu\upsilon}$, a differential equation for the trajectory $x_o^\mu(t)$ of the object:

$$\Gamma^\lambda_{\mu\upsilon}(g(x_O))\frac{dx_O^\mu}{d\tau}\frac{dx_O^\upsilon}{d\tau} + \frac{d^2 x_O^\lambda}{d\tau^2} = 0 \qquad (11)$$

This equation has been used to explain various phenomena such as the precession of planetary orbits, and after adaptation to zero-mass particles, the bending of light by the sun, the gravitational redshift of light, and the time delay of radar signals within the solar system. To that end the metric must be specified, and these key tests have all involved the Schwarzschild metric

$$d\tau^2 = \left(1-\frac{2GM}{c^2 r}\right)dt^2 - \frac{r^2}{c^2}\left(d\theta^2 + \sin^2(\theta)d\varphi^2\right) - \frac{dr^2}{c^2\left(1-\frac{2GM}{c^2 r}\right)} \qquad (12)$$

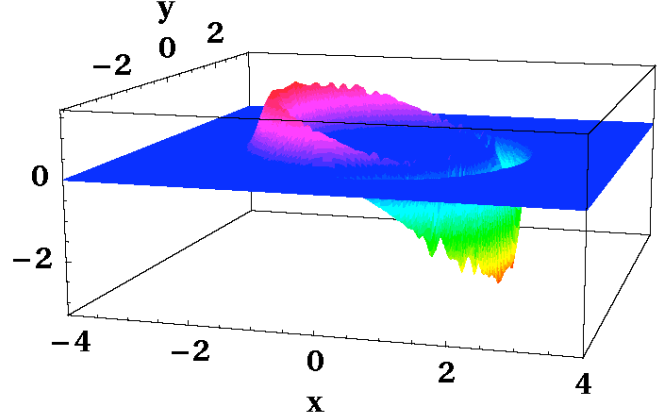

Fig.5 Shows the 'dark matter' density from (7) when the velocity field is given by the propagating bubble in (19)-(20). The plot shows the effective matter density on a plane passing through the centre of the bubble. This effective matter density, which resides in the boundary layer of the bubble, is negative in some regions. However this is physical for the new theory of space as this effective matter density is simply a means of describing the spatial self-interaction dynamics in (5). However we see that for the bubble in (19)-(20), this effective matter density is not the same as required for stability of the bubble, i.e. (19)-(20) does not satisfy (5). But perhaps a modified bubble velocity field may do so. If not then any residual stabilisation effects could be engineered by using ordinary matter and/or electromagnetic fields. If the bubble is evolved in time using (2), but with no matter density, then as shown in Fig.6, shock waves develop from the leading surface, and propagate back into the bubble, leading to its eventual decay.

where $M$ is either the mass of the sun or the earth. This metric is a solution of Einstein equation, external to a spherically symmetric mass, for the metric $g_{\mu\nu}$

$$G_{\mu\nu} \equiv R_{\mu\nu} - \frac{1}{2} R g_{\mu\nu} = \frac{8\pi G}{c^2} T_{\mu\nu} \qquad (13)$$

where the Riemann tensors $R_{\mu\nu}$ and $R$ depend on $g_{\mu\nu}(x)$, and $T_{\mu\nu}$ is the energy-momentum tensor. The standard interpretation of these successful tests is that the metric describes an existing/physical curved four-dimensional manifold; this is the spacetime ontology, and here the curvature is the explanation for the phenomenon of gravity. However we shall now see that the curved spacetime ontology is an incorrect interpretation of (12), but that even more significantly (13) is in conflict with much experimental and observational data. To see this we first make the change of variables, discovered by Panlevé and Gullstrand in the 1920's: $t \rightarrow t'$ and $r \rightarrow r' = r$ with

$$t' = t + \frac{2}{c} \sqrt{\frac{2GMr}{c^2}} - \frac{4GM}{c^2} \tanh^{-1} \sqrt{\frac{2GM}{c^2 r}} \qquad (14)$$

Then the Schwarzschild solution (12) takes the form

$$d\tau^2 = dt'^2 - \frac{1}{c^2} \left( dr' + \sqrt{\frac{2GM}{r'}} dt' \right)^2 - \frac{r'^2}{c^2} \left( d\theta'^2 + \sin^2(\theta') d\varphi'^2 \right) \qquad (15)$$

which is an equally valid description of the spacetime as the interval measure is invariant under a change of coordinate description of the manifold. Eqn.(15)

shows that the external-Schwarschild metric is specified by the in-flowing velocity field in (3). To explore this insight we consider the more general class of 'flow-metrics' of the form

$$d\tau^2 = g_{\mu\nu}dx^\mu dx^\nu = dt^2 - \frac{1}{c^2}(d\mathbf{r} - \mathbf{v}(\mathbf{r},t)dt)^2 \qquad (16)$$

where $\mathbf{v}(\mathbf{r},t)$ is a time-dependent and inhomogeneous velocity field. Then (10) takes the form

$$\tau[\mathbf{r}_O] = \int dt \left(1 - \frac{\mathbf{v}_R^2}{c^2}\right)^{1/2} \qquad (17)$$

where $\mathbf{v}_R = \mathbf{v}_O - \mathbf{v}$, with $\mathbf{v}_O$ the velocity of the object, with position $\mathbf{r}_O(t)$, relative to the local space. Then (11) takes the explicit form

$$\frac{d\mathbf{v}_O}{dt} = \left(\frac{\partial \mathbf{v}}{\partial t} + (\mathbf{v}.\nabla)\mathbf{v}\right) + (\nabla \times \mathbf{v}) \times \mathbf{v}_R - \frac{\mathbf{v}_R}{1 - \frac{\mathbf{v}_R^2}{c^2}} \frac{1}{2} \frac{d}{dt}\left(\frac{\mathbf{v}_R^2}{c^2}\right) \qquad (18)$$

This form is remarkably revealing. The 1st term is the Euler 'fluid' acceleration in (1), the 2nd term is the vorticity-induced Helmholtz acceleration, and the last is a relativistic effect leading to the so-called 'geodesic' effects, such as the precession of elliptical orbits. So the metric (16) reveals a close link between the spatial flow phenomenon and relativistic effects, and (16) includes all the special cases where the spacetime ontology was supposed to have been directly checked. This implies that the Einstein equation (13) may really have been about a velocity flow-field all along. To extract the explicit form of that equation we substitute the metric in (16) into (13) and we arrive exactly at (2) in the non-relativistic limit. This analysis shows that the famous external-Schwarzschild metric is nothing more than Newtonian gravity in disguise, and that the so-called tests of GR were really testing the trajectory equation (18) in which the 'metric' was encoding the in-flow velocity field in (3). Of course by using a more general coordinate system this simple observation has been well hidden. Reformulation of (17) for electromagnetic waves gives the gravitational light bending and in particular the gravitational lensing effect. Hence the experimental evidence is that gravity is really explained by the time-dependence and spatial inhomogeneities of a velocity field. Hence we see that a neo-Lorentzian effect is taking place here; the motion of an object through a 3-space in differential internal relative motion causes accelerations of that object that we know as gravity. This is not merely an interpretation: the internal motion of that space as well as the absolute motion through that space has been detected in several experiments, particularly by Miller [14] and DeWitte [3]. So the *spacetime metric* was really all along describing a spatial-flow phenomenon in those cases where it was supposed to have been tested. Of course there are metrics not equivalent to the form in (16), and for which GR does not reduce to (2) in the absence of vorticity. But these metrics have never been directly tested by experiment or observation and so contribute nothing to the above. There is in fact only *one* indirect confirmation of the GR formalism, apart from the misleading external-Schwarzschild metric cases, namely the observed decay of the binary pulsar orbital motions, for only in this case is the metric non-Schwarzschild, and so not equivalent to the 'inverse square law'. However the new theory of gravity also leads to the decay of orbits, and on the grounds of dimensional analysis we would expect comparable predictions. It is also usually argued that the Global Positioning System (GPS) demonstrated the efficacy of General Relativity. However the new spatial-flow formalism of gravity also explains this system, and indeed gives a physical insight into the processes involved. In particular the relativistic speed and `gravitational red-shift' effects now acquire a unified explanation.

## 4 Propagating Quantum-Foam Bubble Dynamics

In the context of GR Alcubierre's propagating bubble involves the metric of the form in (16) where

$$\mathbf{v}(\mathbf{r},t) = (v_s f(r_s(t)), 0, 0); \quad r_s(t) = \left((x - v_s t)^2 + y^2 + z^2\right)^{1/2} \quad (19)$$

which describes a spherical bubble of space moving with speed $v_s$ in the $+x$ direction, as shown in Fig.2 ($f(r_s(t)) = 1$ at the centre of the bubble), and where the key property is that this speed is not restricted to being less than the speed of light, as it is not matter which has this speed through the space in which it is located. Ordinary matter could indeed be located at the centre of the bubble and so would be at rest with respect to the space in which it is located, but which at the same time would be travelling faster than the speed of light with respect to the external space. The function $f(r)$ models the boundary profile, and Alcubierre chose

$$f(r) = \frac{\tanh(\sigma(r+R)) - \tanh(\sigma(r-R))}{2\tanh(\sigma R)} \quad (20)$$

which gives the bubble a radius $R$ and a surface profile parametrised by $\sigma$. Ignoring the vorticity, so that (2) is the explicit form for the GR spatial bubble dynamics, which is valid if the matter does not have a velocity large compared to $c$, we can compute from (2) the form of the matter density required for the velocity field to satisfy (2); this gives the matter density shown in Fig.4. As is now well known [9,15], but only within the geometrical spacetime formalism of GR, this matter density must be negative in certain sections of the bubble interface, and so would require what is called 'exotic matter'. As well we find that there is a non-zero vorticity, shown in Fig.3, and this would require circulating matter according to (8).

As a part of a preliminary analysis of the Alcubierre bubble dynamics within the new theory of space we can extract using (7) the form of the 'dark matter' density that would have to manifest in order for (19) to be a solution of (5), This gives the 'dark matter' density shown in Fig.5, and this involves regions of negative 'dark matter'; however this is not an exotic form of matter, and merely indicates the nature of the spatial self-interaction dynamics that must take place at the boundary. Comparing Fig.4 and Fig.5 we see that the bubble

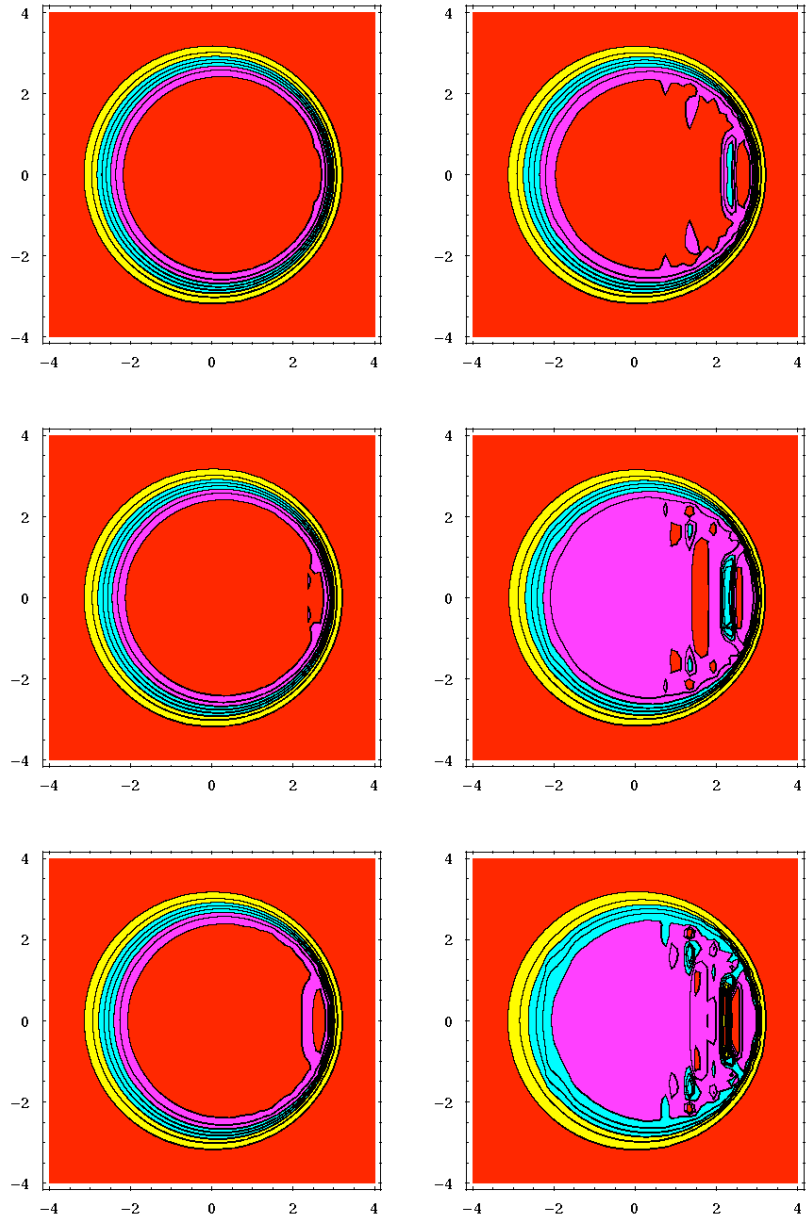

Fig.6 Shows the magnitude of the x component of the velocity field of the propagating bubble in (19)-(20) as it evolves in time according to (2), but with no matter or 'dark matter' density. The x direction is the abscissa, and the bubble is propagating to the right. The section is a plane including the centre of the bubble. The time ordering is via columns, with the earliest time at the top LHS, and the last time at the bottom RHS. As the bubble evolves shock waves develop at the leading

surface, which propagate back into the bubble, resulting eventually in its decay. To dynamically stabilise the bubble, i.e. so that it satisfies (5), a different velocity form from that in (19)-(20) may be successful, or alternatively the stabilisation may be provided by engineering the matter density so that the negative matter density is effectively provided by the spatial self-interaction dynamics in (5).

characterised in (19)-(20) does not satisfy (5) as the required and induced density are not identical. However there may be a modified form for (19) which is a stable propagating bubble solution of (5). To find this form would require either finding analytic solutions to (5) or starting the time evolution in numerical computations with the form in (19)-(20) and evolving that forward in time with (5) to see if a modified stable form emerges. If either of these approaches were successful then we would have a strong case for believing that such faster-than-light bubbles could occur as a natural phenomenon. One intriguing role for these would be in the escape of matter and/or information outwards through the event horizon of black holes. If there are no natural solutions of (5) with any propagating bubble form, then the next stage of investigation is to discover bubble forms which can be stabilised by engineered non-exotic matter and/or electromagnetic fields. This would amount to engineering the quantum foam, and idea that Puthoff [16-18] has discussed in a different context.

The high non-linearity of (5) makes computing numerical solutions difficult. As a first step in this direction the time evolution of the bubble profile in (19)-(20) has been evolved forward in time using (2) with no matter present, either normal or exotic, and so also ignoring vorticity effects. The resulting time evolution of the bubble velocity field is shown in Fig.6. Because there is no matter/'dark matter' present to stabilise the propagating bubble we see that the bubble begins to decay, with 'shock waves' forming at the leading surface which propagate back into the interior of the bubble. Over longer time intervals these waves totally destroy the bubble integrity, and only residual waves survive that carry away the disturbance into the surrounding space.

5 Conclusions

This brief look at the possibility of engineering the quantum foam has raised numerous intriguing possibilities that warrant further detailed investigation. Indeed this would be a quantum-gravity based technology, as the spatial self-interaction dynamics, which is the key to this re-visiting of Alcubierre's warp drive, involves the fine structure constant, suggestive of quantum processes at the deeper levels of the phenomena which we know of as space.